# Yaw Stability Control System Development and Implementation for a Fully Electric Vehicle

Kerim Kahraman*, Mutlu Senturk*, Mümin Tolga Emirler*, Ismail Meriç Can Uygan*
Bilin Aksun-Guvenc*, Levent Guvenc*, Baris Efendioglu**

*Mekar Lab., Okan University Tuzla Campus, Akfırat-Tuzla, İstanbul, Turkey, TR-34959 (e-mail: levent.guvenc@okan.edu.tr).
** Tofaş-Fiat Research and Development, Bursa, Turkey (e-mail: baris.efendioglu@tofas.com.tr)}

Abstract: There is growing interest in fully electric vehicles in the automotive industry as it becomes increasingly more difficult to meet new and upcoming emission regulations based on internal combustion engines. As a result, some auto makers are already introducing a variety of fully electric vehicles into the commercial market while other automotive companies are building and evaluating research prototypes. Fully electric vehicles do not have an internal combustion engine. Hence, drive torque change for a traction control system and for a yaw stability control system has to be through the electric motor used for traction. The regenerative braking capability of fully electric vehicles has to be taken into account in designing braking controllers like ABS and yaw stability control through differential braking. Fully electric vehicles are usually lighter vehicles with different dynamic characteristics than that of their predecessors using internal combustion engines. As such, their yaw stability control systems have to be re-designed and tested. This paper reports the initial results of ongoing work on yaw stability controller design for a fully electric vehicle. Two different implementations on a research prototype fully electric light commercial vehicle are considered. The first implementation uses the production yaw stability control system in the internal combustion engine powered conventional vehicle. The drive torque change commands from the production ECU are read, modified and sent to the electric motor driver in trying to mimic the conventional vehicle. The differential braking commands are the same as in the conventional vehicle. In the second implementation, a generic yaw stability control system that calculates and issues its own drive torque change commands and differential braking commands is designed and implemented. Offline simulations on a validated model and a hardware-in-the-loop simulation system are used in designing the yaw stability control system. The developed control system will be tested on the prototype vehicle after lab testing and development.

*Keywords:* automotive control, fully electric vehicle, yaw stability control.

## 1. INTRODUCTION

The Yaw Stability Control (YSC) system in a road vehicle is an active safety control system that prevents the vehicle from skidding by selectively applying individual brakes and lowering drive torque through the engine management system. If the actual behaviour of the road vehicle differs significantly from its expected yaw dynamic behaviour, calculated from the driver's intent by measuring the steering wheel angle, then the YSC system determines that a critical situation may occur and that intervention is necessary (Van Zanten, 1995).

Generally two types of vehicle instability occur: understeer and oversteer. Understeer occurs when the slip angle of the front tires is greater than that of the rear tires. Thus in order to negotiate the curve, a greater steer angle is required. The vehicle tends to drift outside of the curve during understeer. The oversteer situation occurs if the slip angle of the rear tires is greater than that of the front tires. The vehicle tends to drift towards the inside of the curve in oversteer. Understeer and oversteer both involve a steering input by the driver that the vehicle does not respond to as is desired. It is also possible for a road vehicle to exhibit undesired yaw motion in the absence of driver steering input, due to a yaw moment disturbance. A yaw moment disturbance may be caused by sudden side wind or a µ-split driving situation with uneven road friction characteristics.

If a vehicle experiences an understeer situation, the YSC system activates the rear brake on the inside of the turn and also intervenes in the engine management system by reducing drive torque. When the vehicle tends to oversteer, the YSC system activates the front brake on the outside of the curve and intervenes in the engine management system by sending torque reduction commands.

Hybrid electric and electric vehicles are being more popular as fossil resources are declining and environmental issues are becoming more demanding as engine and exhaust gas emission regulations are becoming stricter. As a result, automotive producers are seeking for vehicles powered by alternative energy sources. Consequently, fully electric vehicles are entering the commercial market in growing numbers and are currently viewed as the future of automobile technology as they do not directly pollute the environment.

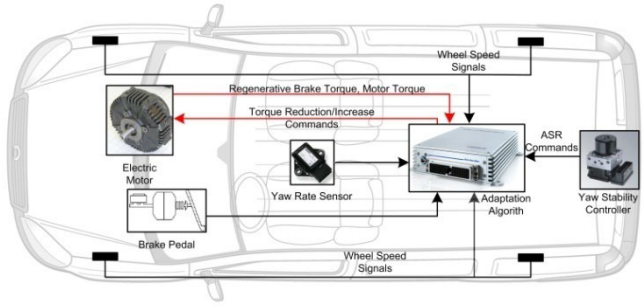

Fig. 1. YSC Adaptation Algorithm Implementation for a Fully Electric Vehicle

In the literature, most of the studies on fully electric vehicle stability control systems are on four wheel drive vehicles using in-wheel motors at the wheels. Four wheel driven systems can provide vehicle stability by using the braking and acceleration abilities of independent in-wheel electric motors on each wheel. Lateral dynamic control based on electric motor torque control has shown better performance than conventional ESP in simulations (Feiqianq, 2009). While being hindered by the battery range problem, electric vehicles also possess several advantages as compared to internal combustion engine vehicles (Hori, 2004).

Regenerative braking is used to generate energy while braking in electric vehicles (Hancock, 2006). Using electric motors for providing vehicle stability has advantages such as faster response time and the capability of using regenerative braking. To make the vehicle more stable, brake intervention may be needed because the wheels on the same axle cannot be controlled independently.

Electric vehicles have some advantages over internal combustion engines from the viewpoint of motion control (Yin, 2009). They are capable of quick and accurate torque generation. The motor torque can be measured easily so that the driving and braking torque can easily be estimated. Their regenerative braking capabilities are useful in the presence of environmental concerns and also recuperation of energy improves fuel economy.

Active safety systems for fully electric vehicles are an important concern. Independent hub electric motor technology is not mature enough for commercial application yet. Thus, the anti-lock brakes (ABS), traction control system (TCS) and yaw stability control system (YSC) have to rely on the conventional brake system, a hydraulic system in this paper. The ABS, TCS and YSC systems have to be re-designed for a fully electric vehicle for successful mass production.

The regenerative braking capability of fully electric vehicles has to be taken into account in designing braking controllers like ABS. Fully electric vehicles are usually lighter vehicles with different dynamic characteristics than that of their predecessors using their internal combustion engines. As such their yaw stability control systems have to be re-designed and re-tested. This paper reports the initial results of ongoing work on yaw stability controller design for a fully electric vehicle. Two different implementations on a research prototype fully electric light commercial vehicle are considered.

The first implementation uses the production yaw stability control system in the internal combustion engine powered conventional predecessor vehicle. Thus, an adaptation algorithm is necessary to satisfy vehicle dynamic stability (see Fig. 1). The YSC can apply the hydraulic brakes individually as in a conventional vehicle. The YSC may also request torque reduction or increase from the drive motor. At this point, YSC torque commands need to be adapted to the characteristics of the drive electric motor. In internal combustion engine driven vehicles, the YSC sends those torque requests to the engine management ECU over the vehicle Controller Area Network (CAN) interface. The engine management ECU can reduce engine torque by closing the throttle valve, reducing the spark advance or inhibiting fuel injection (Schuette, 2005).

In the second implementation, a generic yaw stability control system that calculates and issues its own drive torque change commands and differential braking commands is designed and implemented.

The remainder of this paper is organized as follows. In Section 2, the fully electric vehicle model used is introduced. Single track, double track and higher fidelity CarSim vehicle models are introduced. In Section 3, the results of a model validation study with data taken from the actual prototype fully electric vehicle is presented. Initial results of ongoing work on how the YSC system for drive torque request and differential braking is designed is given in Section 4. Section 5 contains the structure of the Hardware-in-the-Loop (HiL) setup used for development and testing of the YSC system in the lab. The paper ends with conclusions in its last section.

2. FEV MODELING

The three fully electric vehicle models used in this paper are the linear single track model, double track model and the higher fidelity Carsim model.

2.1 Single Track Model

The single track model also known as the bicycle model is the simplest lateral dynamics model that matches actual road vehicle lateral dynamics quite accurately for up to 0.4g of lateral acceleration. This model is linearized and used in YSC system design here. The design is then evaluated using all the models presented in this section in offline simulations before the subsequent HiL and road tests.

The linearized single track vehicle model can be expressed in state space form as (see Ackermann, et al., 2002; Aksun-Guvenc and Guvenc, 2002, Aksun-Guvenc et al, 2003 for example):

$$dx/dt = Ax + Bu, y = Cx \qquad (1)$$

where

$$u = \delta, x = [\beta \quad r]$$

$$A = \begin{bmatrix} \dfrac{-(C_{fo} + C_{ro})\mu}{mV} & -1 + \dfrac{(C_{ro}l_r + C_{fo}l_f)\mu}{mV^2} \\ \dfrac{(C_{ro}l_r - C_{fo}l_f)\mu}{I_z} & \dfrac{-(C_{ro}l_r^2 + C_{fo}l_f^2)\mu}{I_zV} \end{bmatrix}$$

$$B = \begin{bmatrix} \dfrac{C_{fo}\mu}{mV} \\ \dfrac{C_{fo}l_f\mu}{I_z} \end{bmatrix}$$

In equation (1), $\delta = \delta_f$ is the steering angle, $\beta$ is the vehicle side slip angle, $r$ is the yaw rate, $C_{fo}$ and $C_{ro}$ are the front and rear tire cornering stiffness values, $l_f$ and $l_r$ are distances from the front and rear axles to the vehicle center of gravity, $m$ is the vehicle mass, $I_z$ is the yaw moment of inertia, $V$ is the vehicle velocity and $\mu$ is the tire road friction coefficient.

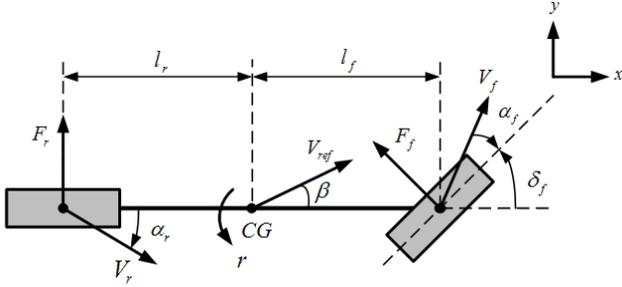

Fig. 2. Illustration of the single track vehicle model.

Fig. 2 shows an illustration of the single track vehicle model. $F_f$ and $F_r$ are the lateral tire forces, $\alpha_f$ and $\alpha_r$ are the tire side slip angles.

2.2 Double Track Model

The double track model is used to test the control algorithm. The double track vehicle model consists of lateral and longitudinal dynamics of a vehicle. Vertical, pitch and roll dynamics are neglected. So this model is the simplest vehicle model with all four wheels.

The geometry of the double track model is displayed in Fig. 3.

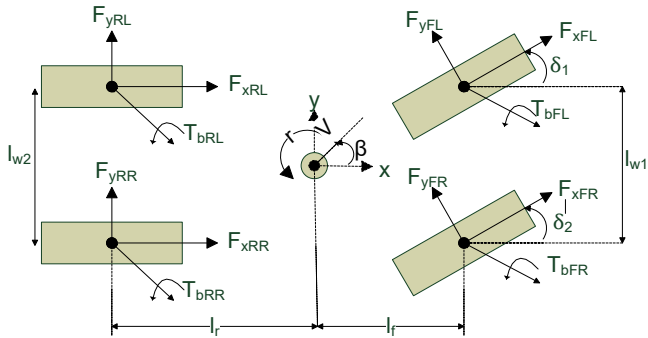

Fig. 3. Illustration of the double track vehicle model.

The force and moment balance dynamic equations of the double track model are given as

$$ma_x = F_{xFL}\cos\delta_1 + F_{xFR}\cos\delta_2 + F_{xRL} + F_{xRR} \\ - F_{yFL}\sin\delta_1 - F_{yFR}\sin\delta_2 \quad (2)$$

$$ma_y = F_{yFL}\cos\delta_1 + F_{yFR}\cos\delta_2 + F_{yRL} + F_{yRR} \\ + F_{xFL}\sin\delta_1 + F_{xFR}\sin\delta_2 \quad (3)$$

$$I_z\dot{r} = [F_{yFL}\cos\delta_1 + F_{yFR}\cos\delta_2 + F_{xFL}\sin\delta_1 \\ + F_{xFR}\sin\delta_2] \\ + [F_{yFL}\sin\delta_1 \\ - F_{yFR}\sin\delta_2 - F_{xFL}\cos\delta_1 \\ + F_{xFR}\cos\delta_2]\dfrac{l_{w1}}{2} \\ + [-F_{yRL} - F_{yRR}]l_r \\ + [-F_{xRL} + F_{xRR}]\dfrac{l_{w2}}{2} \quad (4)$$

The variables used in the double track model are defined in Table 1. The subscript '$i$' defines the tire under consideration as FL, FR, RL, RR.

Table 1. Double track vehicle parameters

| Symbol | Definition | Unit |
|---|---|---|
| $a_x$ | Vehicle longitudinal acceleration | m/s² |
| $a_y$ | Vehicle lateral acceleration | m/s² |
| $r$ | Yaw rate | rad/s |
| $I_w$ | Wheel moment of inertia | kg.m² |
| $w_i$ | Wheel rotational velocity | rad/s |
| $T_d$ | Drive torque on wheels | N.m |
| $T_{bi}$ | Braking torque | N.m |
| $F_{xi}$ | Tire longitudinal velocity | N |
| $l_{w1}, l_{w2}$ | Front and rear track width | m |
| $R_w$ | Effective tire radius | m |
| $F_{ri}$ | Tire rolling resistance force | N |
| $V_{xi}$ | Wheel longitudinal velocity | m/s |
| $V_{wi}$ | Wheel center of gravity velocity | m/s |
| $\alpha_i$ | Wheel side slip angle | rad |
| $V_x$ | Vehicle longitudinal velocity | m/s |
| $V_y$ | Vehicle lateral velocity | m/s |
| $\lambda_i$ | Wheel slip ratio | - |

The tire side slip angles are:

$$\alpha_{FL} = \delta - \tan^{-1}\left(\dfrac{V_y + l_f r}{V_x - \dfrac{l_{w1}}{2}r}\right) \quad (5)$$

$$\alpha_{FR} = \delta - \tan^{-1}\left(\dfrac{V_y + l_f r}{V_x + \dfrac{l_{w1}}{2}r}\right) \quad (6)$$

$$\alpha_{RL} = -\tan^{-1}\left(\dfrac{V_y - l_r r}{V_x - \dfrac{l_{w2}}{2}r}\right) \quad (7)$$

$$\alpha_{RR} = -\tan^{-1}\left(\dfrac{V_y - l_r r}{V_x + \dfrac{l_{w2}}{2}r}\right) \quad (8)$$

where $\delta = \delta_1 = \delta_2$.

The longitudinal wheel slip ratio for each tire is given by,

$$\lambda_i = \begin{cases} \dfrac{\omega_i R_w - V_{xi}}{V_{xi}}, \omega_i R_w - V_{xi} < 0 \\ \dfrac{\omega_i R_w - V_{xi}}{\omega_i R_w}, \omega_i R_w - V_{xi} \geq 0 \end{cases} \quad (9)$$

2.3 CarSim Model

CarSim is a commercial program and provides a highly realistic, high fidelity vehicle model. The Carsim model includes a driver model (if needed), longitudinal, lateral and vertical dynamics and is real time capable.

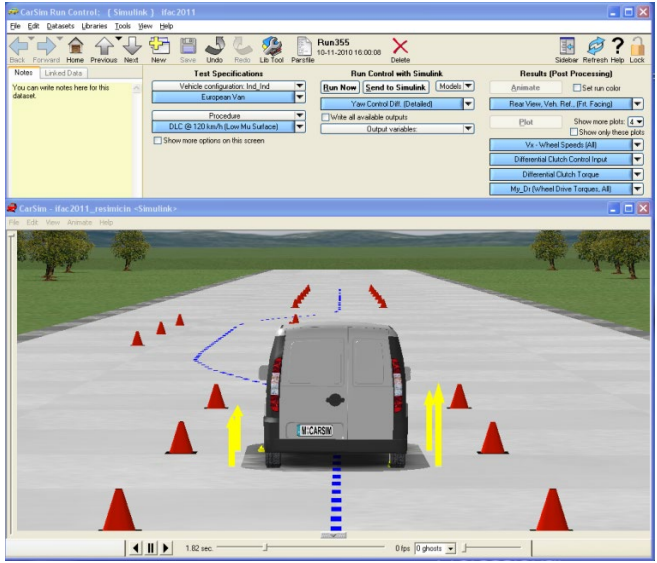

Fig. 4. CarSim Model Screen

The fully electric light commercial vehicle considered in this paper is modelled and its YSC algorithm is tested by using the CarSim vehicle model in offline and HiL tests before proceeding to road testing. Fig. 4 shows an animation screenshot of the Carsim vehicle model in a double lane change maneuver.

2.4 Longitudinal Dynamics

The longitudinal dynamics model of the system shown in Fig. 5 is given by

$$Ma_x = F_b + (F_{aero} + F_{ri} \pm F_{hc}), \ddot{x} < 0 \quad (10)$$

The last three terms on the right hand side of (10) represent the road load whereas the first term represents the tire longitudinal force component generated due to braking. Therefore, the amount of deceleration depends on the vehicle mass, the amount of brake applied and the resistive forces such as aerodynamic drag ($F_{aero}$), rolling resistance ($F_{ri}$) and road slope resistance ($F_{hc}$) that act upon the vehicle. $F_{hc}$ may be negative or positive depending on uphill or downhill motion, respectively.

The powertrain model also shown in Fig. 5 consists of the battery, the electric motor model and the wheel dynamics model.

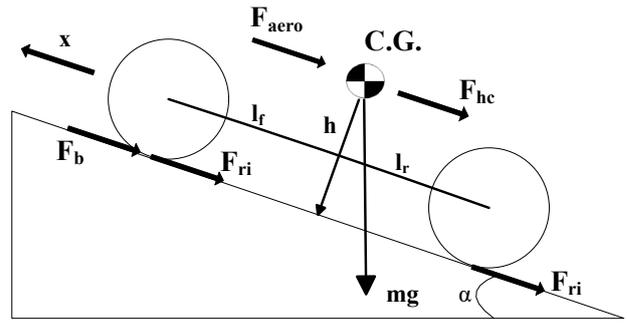

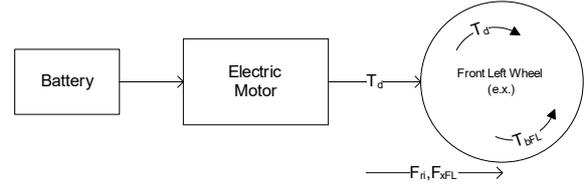

Fig. 5. Longitudinal Dynamics and Powertrain Models (One Wheel Only)

The wheel dynamics can be represented as

$$I_{wi}\dot{\omega} = T_d - T_{bi} - F_{xi}R_w - F_{ri}R_w, \quad (11)$$

where i=FL,FR,RL,RR.
The electric motor model can be expressed as

$$V_q = R_q i_q + L_q \dfrac{di_q}{dt} + K_b w_m \quad (12)$$
$$T_m = K_t i_q \quad (13)$$

where $V_q$ is the voltage source, $i_q$ is the motor current, $R_q$ is the motor resistance, $L_q$ is the motor inductance, $K_b$ is the back e.m.f. constant, $w_m$ is the motor speed, $T_m$ is the motor torque, $K_t$ is the torque constant.

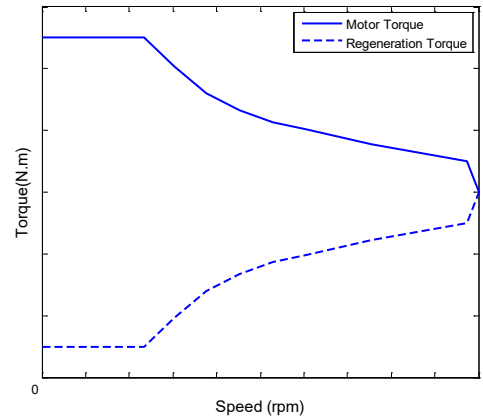

Fig. 6. Electric Motor Speed-Torque Map

The drive torque acting on the wheels is

$$T_d = T_m \eta_t k \quad (14)$$

where $\eta_t$ is the transmission efficiency and $k$ is the transmission gear ratio. In this study, the experimental speed-torque characteristic shown in Fig.6 is used for both acceleration and the regenerative braking.

## 2.5 Tire Models

Two tire models are used for the vehicle models. A linear tire model is used for the linear single track model. Tire force characteristics are linearized as

$$F_f = \mu C_{fo} a_f \tag{15}$$
$$F_r = \mu C_{ro} a_r \tag{16}$$

where

$$a_f = \delta_f - \left(\beta + \frac{l_f r}{V}\right) \tag{17}$$
$$a_r = -\left(\beta - \frac{l_r r}{V}\right) \tag{18}$$

A modified Pacejka tire model that accurately represents tire force saturation and the coupling between longitudinal and lateral tire forces is used in the double track and CarSim vehicle models (Pacejka, 2002). Validated Pacejka tire model parameters were provided by Tofaş-Fiat.

The simpler Pacejka tire model equations that do not involve coupling are

$$F_x = D \sin\left[C \tan^{-1}\left(B(\lambda + S_h)\right.\right. \\ \left.\left. - E(B(\lambda + S_h)\right.\right. \\ \left.\left. - \tan^{-1}(B(\lambda + S_h)))\right)\right] + S_v \tag{19}$$

$$F_y = D \sin\left[C \tan^{-1}\left(B(\alpha + S_h)\right.\right. \\ \left.\left. - E(B\alpha - \tan^{-1}(B( + S_h)))\right)\right] + S_v \tag{20}$$

where $B, C, D, E, S_h, S_v$ are determined by experimental tests and have no physical meaning.

## 3. MODEL VALIDATION STUDY

In this section, the double track fully electric vehicle model is validated by using the experimental vehicle data provided by Tofaş-Fiat in Turkey based on experiments run using their first generation research prototype fully electric vehicle.

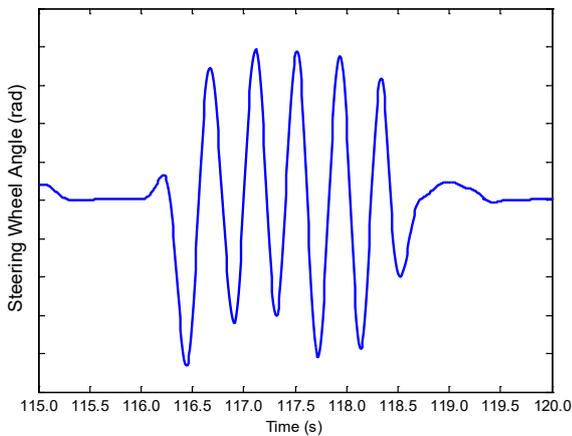

Fig. 7. Experimental Vehicle Steering Angle

The steering wheel angle input in the experiment under consideration is shown in Fig. 7. Fig.8 shows the corresponding actual and double track vehicle model yaw rate responses at about 85 km/h of vehicle speed.

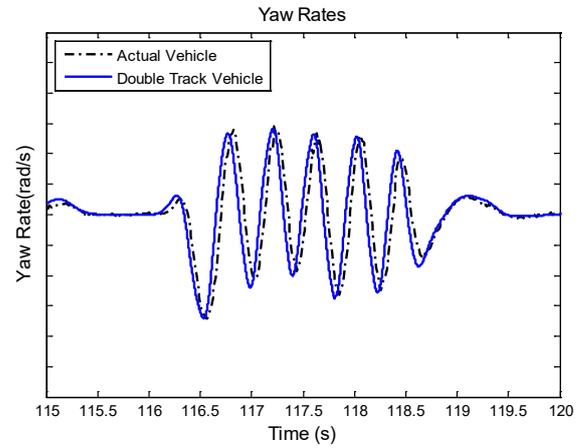

Fig. 8. Experimental Vehicle Yaw Rate Response

Both the actual vehicle and the double track vehicle model have their yaw stability control system function turned on during this maneuver. The double track vehicle response is quite similar to that of the actual vehicle except for a small phase difference.

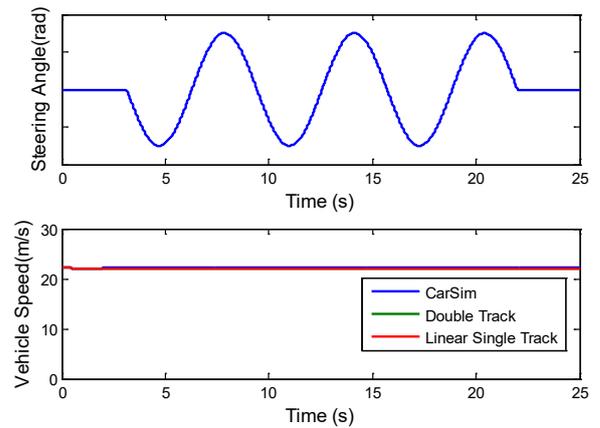

Fig. 9. Steering Angles and Vehicle Speeds for Models

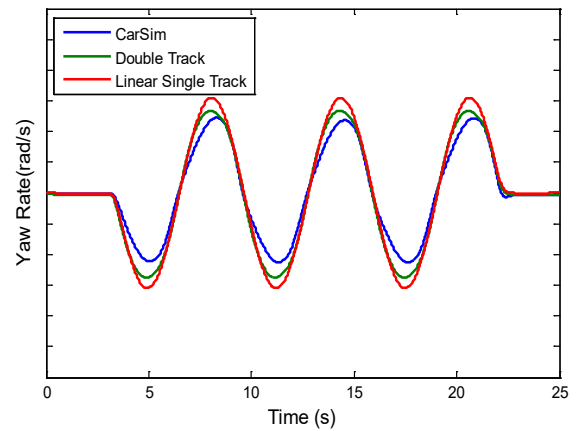

Fig. 10. Yaw Rate Responses of the Models

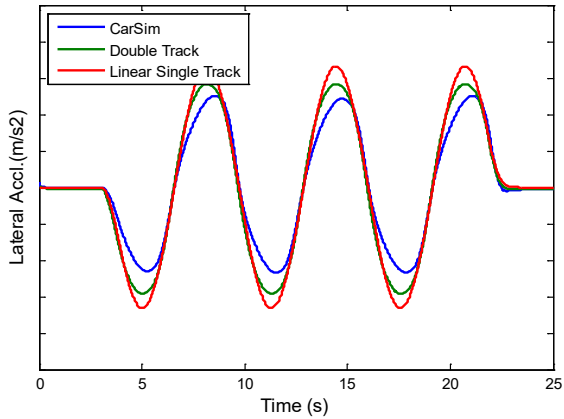

Fig. 11. Lateral Acceleration Responses of the Models

Fig.9-11 shows comparison plots between CarSim, double track and linear single track models. According to these plots, the three models give similar responses.

4. YSC (YAW STABILITY CONTROL) SYSTEM

YSC systems are manufactured for internal combustion engine powered conventional vehicles. The simplest approach in YSC development for a fully electric vehicle will be to adapt the existing YSC system to a fully electric vehicle. In such an adaptation procedure, several modifications should be made. The most important modification is to convert the YSC system engine torque intervention commands to appropriate drive electric motor torque intervention commands and send them to the electric motor driver over the CAN line. Another solution is to design a complete yaw stability control system for a fully electric vehicle. Both of these approaches are treated in this section

4.1 Simple YSC Adaptation through ASR Drive Torque Request Modifications

The yaw stability controller sends ASR Torque commands to the engine when intervention is needed. In a fully electric vehicle however those signals are not meaningful as there is no internal combustion engine and no engine management ECU. Thus, these commands should be translated to commands that are meaningful for the electric motor driver. ASR intervenes in two ways. If only one wheel is spinning, it applies brake at that wheel. This way torque is transferred to the non-spinning wheel. If both traction wheels are spinning ASR request an engine torque reduction. ASR is active when speed differences of the front and rear wheels on the same side are more than 5-6 km/h. When the driver requests more torque from the engine, traction forces cannot transferred to the road. The system sets torque values ASR.TorqueReductionFast or ASR.TorqueReductSlow which are expressed by positive percentage values. The engine torque tries to follow this torque value when ASR is active. In ICE powered vehicles this is achieved by retarding the ignition timing or closing the throttle valve. For an electric motor driven vehicle as motor torque is a function of motor current, motor current must be controlled to achieve the requested torque change.

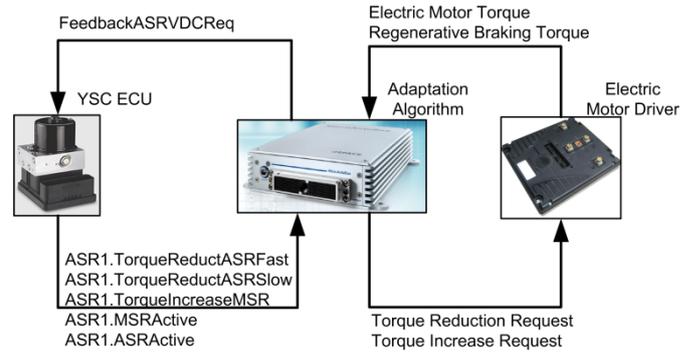

Fig. 12. Implementation of the Adaptation Algorithm

For this purpose, an adaptation of these commands is carried out. ASR signals enter the algorithm as inputs and they are converted to torque increase and decrease signals for the electric motor. The electric motor torque is fed back to the algorithm and intervention status is sent back to the YSC ECU.

4.2 General YSC Algorithm Development and Implementation

The general YSC system consists of two part which are the supervisory upper level controller with decision making and calculations and the lower level controller used for individual braking and electric motor torque reduction command generation operation. The overall controller structure is shown in Fig. 13. While the upper level supervisory controller determines first if the YSC will be activated and if activated whether or not to use electric motor torque intervention and the amount of individual wheel braking to be used, the lower level controller distributes the nominal corrective yaw moment calculated to the brake disks. The lower level controller determines which wheel should be braked and computes hydraulic valve control signals to generate the required brake pressure.

The upper level controller determines the electric motor torque requests and the required friction braking torque. Friction braking torque is controlled with state variable feedback control at present. Other control methods for the individual braking control implementation are also being explored. For the vehicle lateral stability both the vehicle slip angle and yaw rate are so important and should both be controlled. Vehicle yaw rate is measured by a gyroscopic sensor but slip angle sensors are too expensive. The slip angle is estimated with a linear Kalman filter here. Other estimation techniques will be used in later work.

The upper level controller monitors the yaw rate and the slip angle errors by comparing their measured and estimated values with their nominal values. These nominal values can be computed by using the static single track model (Rajamani, 2006), (Emirler et al, 2013).

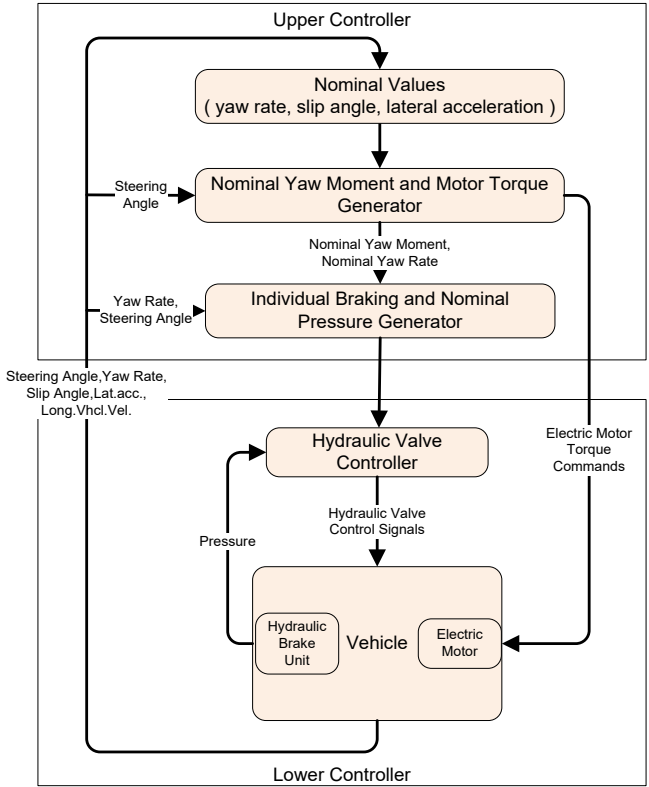

Fig. 13. Controller Structure

$$r_{nom} = \frac{V_x}{l_f + l_r + \frac{mV_x^2(l_r C_{ro} - l_f C_{fo})}{2C_{fo}C_{ro}(l_f + l_r)}} \delta \quad (21)$$

In (21), $V_x$ is the vehicle longitudinal velocity. If the road is wet or snowy, the tires cannot produce high forces for high yaw rates (Rajamani, 2006). Therefore, the target yaw rate must be limited according to the tire-road friction coefficient as represented in (22).

$$r_{limit} = 0.85 \frac{\mu g}{V_x} \quad (22)$$

The vehicle nominal slip angle value is computed similarly as

$$\beta_{nom} = \frac{l_r - \frac{l_f m V_x^2}{2C_{ro}(l_f + l_r)}}{l_f + l_r + \frac{mV_x^2(l_r C_{ro} - l_f C_{fo})}{2C_{fo}C_{ro}(l_f + l_r)}} \delta \quad (23)$$

The target slip angle value must also be limited according to the tire-road friction coefficient as

$$\beta_{limit} = \tan^{-1}(0.02 \mu g) \quad (24)$$

The corrective yaw moment is calculated by a state variable feedback control algorithm. The linear single track model is used for the vehicle model to design the controller.

The control law is:

$$M = Ke \quad (25)$$

where $M$ is the corrective yaw moment, $K$ is the state variable feedback controller gain vector and the $e$ is the vector of state errors between the nominal values and the actual values for the vehicle side slip angle and the yaw rate.

The lower level controller is a rule-based control algorithm that is based on individual wheel braking. After calculation of the required corrective yaw moment by the upper level controller, the lower level controller takes over, computes the individual wheel braking torques that are necessary and applies the corresponding pressure to the appropriate wheel (Emirler, 2009).

The required individual braking moment for each wheel ($T_{bi}$, i = FL, FR, RL, RR) can be calculated using (26) and (27) for the vehicle in Fig. 3 (Falcone, P., 2008) as

$$T_{bFL} = T_{bFR} = \frac{|M|R_w}{\sin\left[\tan^{-1}(\frac{l_{w1}}{2l_f}) - \delta\right] \sqrt{l_f^2 + \left(\frac{l_{w1}}{2}\right)^2}} \quad (26)$$

$$T_{bRL} = T_{bRR} = \frac{|M|R_w}{\sin\left[\tan^{-1}(\frac{l_{w2}}{2l_r})\right] \sqrt{l_r^2 + \left(\frac{l_{w2}}{2}\right)^2}} \quad (27)$$

The main differences of the equations for front and rear wheels arise from the existence of the front wheel steering angle.

In the algorithm development process, all possible cases of vehicle yaw rate error are investigated first. Then, in accordance with these cases, the wheel to be braked is determined. The six cases considered here (Yang, 2009) are tabulated in Table 2. In Table 2, the first column is the case number and the second column states the measured vehicle yaw rate for that case. The desired vehicle yaw rate value is listed in the third column and the fourth column is devoted to indicating the vehicle instant yaw rate situation in comparison with the vehicle's desired yaw rate.

**Table 2. Vehicle cases for lower controller design**

| Case | Vehicle yaw rate | Desired vehicle yaw rate | Situation | Braking wheel |
|---|---|---|---|---|
| 1 | $r > 0$ | $r_{nom} \geq 0$ | $r_{nom} < r$ | Front Left |
| 2 | $r \geq 0$ | $r_{nom} > 0$ | $r_{nom} > r$ | Rear Right |
| 3 | $r < 0$ | $r_{nom} \geq 0$ | $r_{nom} > r$ | Front Right |
| 4 | $r > 0$ | $r_{nom} < 0$ | $r_{nom} < r$ | Front Left |
| 5 | $r \leq 0$ | $r_{nom} < 0$ | $r_{nom} < r$ | Right Left |
| 6 | $r < 0$ | $r_{nom} < 0$ | $r_{nom} > r$ | Front Right |

This table is used in the simulation results of Fig. 14-15 which are based on the double track fully electric vehicle model.

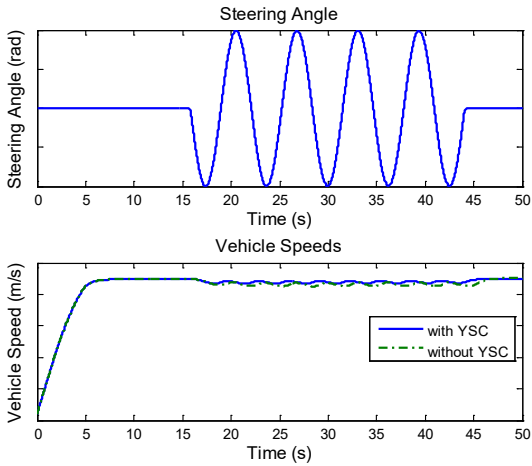

Fig. 14. Vehicle steering angle and Vehicle Speed

The steering angle input and the resulting vehicle speed with and without YSC are shown in Fig. 14. In the simulation, the road is dry and the vehicle speed is about 80 km/h.

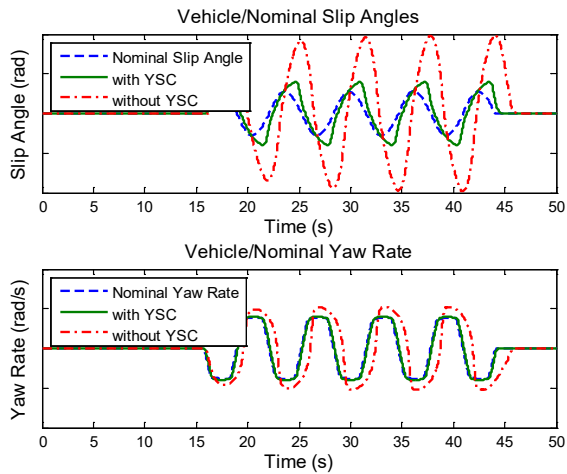

Fig. 15. Simulation Results

Fig.15 shows that the vehicle with YSC tracks the desired nominal slip and yaw rate values much better than the vehicle without YSC.

## 5. HiL SIMULATOR

A Hardware-in-the Loop (HiL) simulator setup is formed to develop and test the YSC systems being designed before final road testing. To activate the YSC in the conventional production ECU all its necessary input signal are generated in real time in the Carsim software and fed to the YSC ECU over the CAN bus and through direct connection of the wheel speed sensor signals.

The production YSC unit consist of the ECU (Electronic Control Unit) and HCU (Hydraulic Control Unit) integrated together. For the HiL tests, the hydraulic unit is separated from the ECU and all its hydraulic parts (solenoid activated valves, return pump, etc.) are modelled in Simulink.

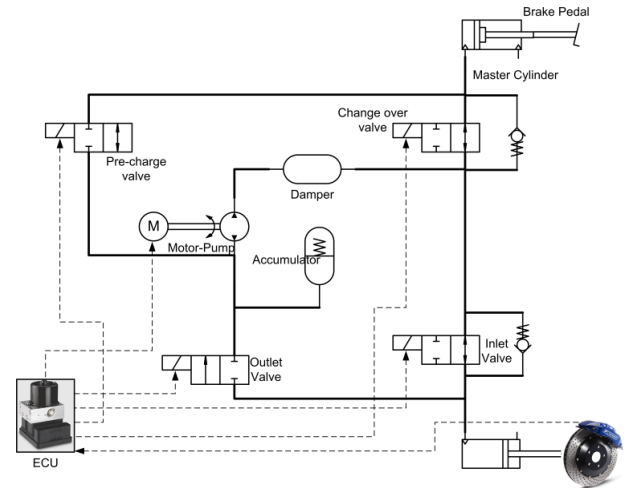

Fig. 16. Brake Hydraulics of YSC for One Wheel

In Fig. 16, the brake system components are shown for just one wheel. When the driver presses the brake pedal, pressure is generated in the wheel brakes. In case of YSC intervention even when the driver is not pushing the brake pedal, hydraulic fluid is sent to the relevant wheel. This is done by activating the hydraulic motor pump and several valves. After pressure is built up, the wheel pressure can be regulated by means of inlet and outlet valves.

The high fidelity, real time CarSim vehicle model and the brake hydraulic model are built and embedded into a dSPACE board. The model computes and delivers all sensor signals like the sensors in a real car. A dSPACE EcoLine Simulator which is capable of generating all the necessary signals like steering wheel angle, yaw rate, lateral and longitudinal accelerations, engine rpm, etc is used as the main HiL setup. The system has a CAN interface to send those signals to the YSC ECU. Then, the controller determines if the soft vehicle is experiencing a critical situation like understeer, oversteer or undesired yaw rate during straight line driving. The YSC activates the necessary magnetic valves to compensate for the undesired yaw motion. To capture the valve signal information a Valve Signal Detection (VSD) unit is used. This unit measures the activity of the magnetic coils which operate the valves in the hydraulic unit. The VSD unit sends the captured solenoid signals to the dSPACE HiL Simulator, where they are used as inputs for the brake hydraulics model. Inside the brake hydraulics model, brake pressures are computed and sent to the CarSim vehicle model. Fig. 17 illustrates the operation of the HiL setup.

## 6. CONCLUSIONS

Intial results of ongoing work on developing a YSC system for a fully electric light commercial vehicle were presented here. This paper gave the details of the development environment along with some preliminary results. The YSC system development used two implementation approaches. The first and simpler one was adapting the YSC ECU in a similar internal combustion engine powered vehicle by monitoring the ICE torque change commands from the YSC system and converting them to equivalent commands for the electric motor driver. In the second, more general implementation method, a generic YSC system that uses its own algorithm for both drive torque reduction and individual wheel braking was developed. The use of a HiL setup for YSC system development within a lab environment was also presented as a method of identifying and solving potential problems before road testing. HiL and road test results will be added to the final version of the paper.

Other approaches like model regulation also called disturbance observer control (Oncu et al, 2007; Aksun-Guvenc and Guvenc, 2002; Guvenc and Srinivasan, 1995), speed scheduled LQR control (Emirler et al, 2015), intelligent control (Boyali and Guvenc, 2010), parameter space based robust control (Guvenc et al, 2017; Emirler et al, 2014, Emirler et al, 2015; Wang et al, 2018) and repetitive control (Demirel and Guvenc, 2010; Necipoglu et al 2011; Orun et al, 2009) for periodic speed profiles can also be applied for yaw stability control and for electric vehicle traction control. The yaw stability controller designed can also be tested as part of an autonomous driving system in an autonomous vehicle hardware-in-the-loop simulator (Gelbal et al, 2017).

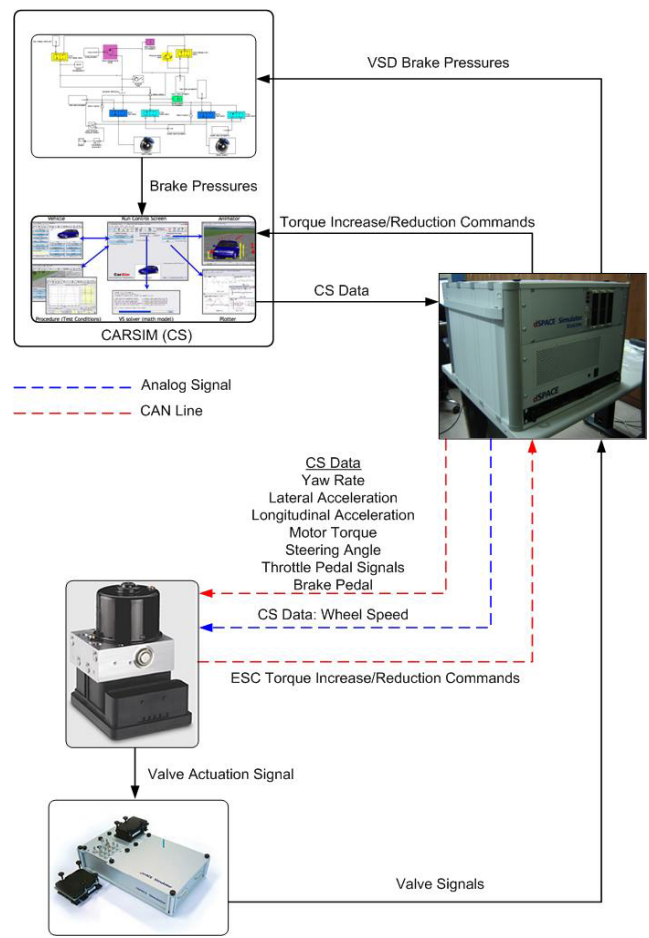

Fig. 17. Hardware In The Loop Simulator Schematic

ACKNOWLEDGMENT

The authors acknowledge the support of Tofaş-Fiat in the work presented here.